\documentclass[12pt,a4paper]{article}

\usepackage{color,tikz}
\usepackage[unicode,bookmarks,bookmarksopen,bookmarksopenlevel=2,colorlinks,linkcolor=blue,citecolor=green]{hyperref}

\usepackage{amsmath,eucal,amssymb,amsthm,amsfonts}
\usepackage{mathrsfs,graphicx,texdraw}

\newtheorem{remark}{Remark}
\DeclareMathOperator{\tr}{tr}

\DeclareMathOperator{\diag}{diag}

\def\neprod{\setbox0=\hbox{$\nearrow$}
  \box0\kern-1.6em\prod} 
\def\swprod{\setbox0=\hbox{$\swarrow$}%
  \,\,\raise.03em\box0\kern-1.18em\prod} 
\def\seprod{\setbox0=\hbox{$\searrow$}%
  \,\,\raise.00em\box0\kern-2.0em\prod} 

\def\im{{\mbox{Im}}}

\def\ad{\mathrm{ad\,}}
\def\openone{\leavevmode\hbox{\small1\kern-3.3pt\normalsize1}}

\def\biglb{\big[\hspace*{-.7mm}\big[}
\def\bigrb{\big]\hspace*{-.7mm}\big]}

\def\diag{\mbox{diag}\,}
\def\tr{\mbox{tr}\,}

\def\fr#1{{\mathfrak{#1}}}

\begin{document}

\begin{center}
{\LARGE \bf On the $N$-waves hierarchy with constant \\[6pt] boundary conditions. Spectral properties}

\bigskip

{\bf Vladimir S. Gerdjikov$^{a,b,}$\footnote{E-mail: {\tt gerjikov@inrne.bas.bg}} and
Georgi  G. Grahovski$^{c,}$\footnote{E-mail: {\tt grah@essex.ac.uk}} }

\end{center}

\medskip

\noindent
{\it $^{a}$ Varna Free University ``Chernorizec Hrabar'',
Chayka Resort, Varna   9007, Bulgaria }\\[5pt]
{\it $^{b}$ Institute of Nuclear Research and Nuclear Energy, Bulgarian Academy of Sciences, \\ 72 Tsarigradsko chausee, Sofia 1784, Bulgaria}\\[5pt]
{\it $^{c}$ School of Mathematics, Statistics and Actuarial Science, University of Essex,\\ Wivenhoe Park,
Colchester CO4 3SQ, UK}

\begin{abstract}
\noindent
The paper is devoted to $N$-wave equations with constant boundary conditions related to symplectic Lie algebras. We study the spectral properties of a class of Lax operators $L$, whose
potentials $Q(x,t)$ tend to constants $Q_\pm$ for $x\to \pm \infty$.
For special choices of $Q_\pm$ we outline the spectral properties of $L$, the direct scattering transform and
construct its fundamental analytic solutions. We generalise
Wronskian relations for the case of CBC -- this allows us to analyse the mapping
between the scattering data and the $x$-derivative of the potential $Q_x$.
Next, using the Wronskian relations we derive the dispersion laws for the
$N$-wave hierarchy and describe the NLEE related to the given Lax operator.

\end{abstract}

\section{Introduction}

The three-wave resonant interactions equations are one of the most important nonlinear models that appeared
at the early stages of development of the inverse scattering method (ISM).  They have numerous applications in physics and have attracted the attention of the
scientific community over the last few decades. Such models describe a special class of wave-wave interactions in nonlinear media, where the nonlinear
dynamics is regarded as a perturbation of linear wave propagation. This explains why they find numerous applications in physics, from plasma physics (radio frequency heating, laser-plasma interactions, plasma instabilities), fluid dynamics
(water waves' interaction), acoustics (light-acoustic interactions), nonlinear optics (frequency conversion, parametric amplification,
 and stimulated Raman scattering) to quantum field theory (the Lee model of weak interactions) and solid state physics (Brillouin scattering).

The $N$-wave resonant interactions model \cite{ZM,K,ZMNP,FaTa,KRB}
\begin{eqnarray}\label{eq:N-waves}
iQ_t - i\, {\rm ad}_J^{-1}{\rm ad}_K\,Q_x +[{\rm ad}_J^{-1}\, {\rm ad}_K\, Q,Q] = 0,\qquad Q=Q(x,t)
\end{eqnarray}
are solvable by the inverse scattering method  \cite{ZM,ZMNP,FaTa}
applied to the generalized system of Zakharov--Shabat type
\cite{ZMNP,Sha1,G*86}:
\begin{eqnarray}\label{eq:L}
&&L(\lambda )\Psi(x,t,\lambda ) = \left(i{d \over dx} + {\rm ad}_J\, Q(x,t)
- \lambda J \right)\Psi(x,t,\lambda ) = 0, \quad J\in \fr{h},\\
\label{eq:1.3.1}
&& Q(x,t)= \sum_{\alpha \in\Delta } Q_\alpha (x,t) E_\alpha \equiv
\sum_{\alpha \in \Delta _+} (q_{\alpha }(x,t)
E_{\alpha } + s_{\alpha}(x,t) E_{-\alpha })
\in \fr{g \backslash \fr{h}}.
\end{eqnarray}
Here the matrices $Q(x,t) $ and $J $ take values in a simple Lie algebra $\fr{g}$, $ \fr{h}$ is the Cartan subalgebra of $\fr{g} $, and $\Delta  $
(resp. $\Delta _+ $) is the system of roots (resp., the system of positive
roots) of $\fr{g} $, while  $E_{\alpha } $ are the root vectors (Weyl generators) of  $\fr{g}$ \cite{Helg}. The adjoint action operator ${\rm ad}_J$ is defined as ${\rm ad}_J\,X=[J,X]$ for $X\in {\fr g}$. Here also $[\cdot,\cdot]$ stays for a standard matrix commutator, and $\lambda$ is a spectral parameter.

One can always choose a representation of $\fr{g}$ where the Cartan subalgebra is represented by diagonal matrices:
\[
\quad J = \sum_{k=1}^{r} a_kH_k, \qquad \quad K = \sum_{k=1}^{r} b_kH_k,
\]
where $H_k$ are the Cartan generators of $\fr{g}$, $r$ is the rank of $\fr{g}$. We will use also the vectors ${\bf a}, {\bf b}\in {\Bbb E}^r$ dual to $J,K\in \fr{h}$, respectively. Here and below we will also assume  that the eigenvalues of $J$ and $K$ are real and the ordering $a_1 >a_2 > \cdots >a_r$. The set $\{H_k,E_{\pm \alpha}\}$, $\alpha \in \Delta_+$ and $k=1,\dots r$ form the Cartan-Weyl basis of $\fr{g}$ (see, e.g. \cite{Helg}).

The $N$-wave equation model \eqref{eq:N-waves} allows soliton solutions. Based on the inverse scattering formalism and the resulting
Riemann-Hilbert problems \cite{Gakhov}, various types of soliton solutions and their interactions were studied in \cite{ZYW,GLX}.

The equation (\ref{eq:N-waves}) is the
compatibility condition
\begin{equation}\label{eq:LM}
[L(\lambda ),M(\lambda )] = 0,
\end{equation}
where
\begin{eqnarray}\label{eq:Mop}
M(\lambda )\Psi(x,t,\lambda ) = \left( i{d \over dt} + {\rm ad}_K\, Q(x,t) -
\lambda K\right) \Psi(x,t,\lambda ) = 0, \quad K \in \fr{h} .
\end{eqnarray}

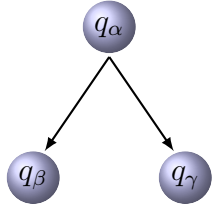
\begin{figure}
\begin{center}
\begin{tikzpicture}
\shade[ball color=blue!20] (0,2) circle (.35cm);
\shade[ball color=blue!20] (-1,0) circle (.35cm);
\shade[ball color=blue!20] (1,0) circle (.35cm);
\draw (0,2) node {$q_\alpha$};
\draw (-1,0) node {$q_\beta$};
\draw (1,0) node {$q_\gamma$};
\draw[thick,-latex] (0,1.6)--(-0.85,0.35);
\draw[thick,-latex] (0,1.6)--(0.85,0.35);
\end{tikzpicture}
\end{center}
  \caption{\small The elementary wave-decay corresponding to each term $H(\alpha ,\beta ,\gamma )$ in \eqref{eq:decay}. If we assign to each wave $q_\alpha$ ($\alpha\in \Delta_+$) a wave number $k_{\alpha } $ and frequency $\omega
_{\alpha }$, then  the resonance condition $\omega (k_{\alpha })=\omega (k_{\beta }) + \omega (k_{\gamma })$ will imply preservation of the wave number: $k_{\alpha } = k_{\beta } + k_{\gamma }$ for each elementary decay $q_\alpha \rightarrow q_\beta + q_\gamma$ (for any triple of roots $\alpha, \beta, \gamma \in \Delta_+$, such that $\beta + \gamma=\alpha$).}\label{fig:decay}
\end{figure}

\noindent
Following the AKNS approach \cite{AKNS,AblSe}, if we take the $M$-operator to be a polynomial function of $\lambda$, then the compatibility condition $[L(\lambda),M(\lambda)]=0$ will give a family (hierarchy) of integrable PDEs, having the same Lax $L$-operator \eqref{eq:L}. The order of the polynomial with respect to $\lambda$ in the $M$-operator determines the order of the resulting PDE. All PDEs from the hierarchy, associated with $L$ are integrable by the ISM. We will briefly outline the next member of the $N$-wave hierarch in  \ref{sec:A1}.

All equations for the hierarchy associated to $L$ \eqref{eq:L} (i.e. the $N$-wave equations and their higher analogues), together with appropriate boundary conditions,
can be solved via the inverse scattering method (ISM). The purpose of this paper is to study equations of the type \eqref{eq:N-waves} in the case when $Q(x,t)$ tends to some
constant matrices $Q_\pm$ for $x\to \pm \infty$. We will assume that $Q(x,t) $  is smooth matrix-valued function for all $x$ and $t$ and is such that
\[
\lim_{x\to\pm\infty} |x|^p (Q(x,t) -Q_\pm) =0\quad  \mbox{for all} \quad p=0,1,\dots.
\]
The interpretation of the ISM as a generalized Fourier transform allows one  to prove that all $N $-wave type equations
are Hamiltonian and possess a hierarchy of pairwise compatible
Hamiltonian structures \cite{gk78,GK*81,G*86,vgrn,vgn1,vgn2,vgn3,Chiu} $\{H^{(k)}, \Omega ^{(k)}\} $,
$k=0,\pm 1,\pm 2, \dots$. The $N$-wave hierarchy associated to \eqref{eq:L}
can be formally written Hamiltonian system $\Omega^{(k)} (Q_t,\cdot ) = dH^{(k)} (\cdot)$, $k=0, \pm1, \pm 2,\dots$,
where the symplectic 1-form $\Omega ^{(k)} $ and the corresponding Hamiltonian $H^{(k)} $ are complex-valued functions. The
simplest (canonical) Hamiltonian formulation of (\ref{eq:N-waves}) is given by
$\{H^{(0)} $, $\Omega^{(0)}\}$, where $H^{(0)}=H_{\rm kin} + H_{\rm int} $ with
\begin{eqnarray}
H_{\rm kin} &=&{1 \over 2i}\int_{-\infty }^{\infty } \, dx\, \left\langle
Q, {\rm ad}_K\, {\rm ad}_J^{-1}\,Q_x \right\rangle = i\sum_{\alpha \in \Delta _+} \int_{-\infty }^{\infty }dx\, {({\bf b},\alpha )\over
({\bf a} ,\alpha )} (q_{\alpha }s_{\alpha ,x} - q_{\alpha ,x}s_{\alpha }),
\\
\label{eq:H-i}
H_{\rm int} &=& {1\over 3} \int_{-\infty }^{\infty }\, dx \, \left\langle
[J,Q],[Q,[K,Q]] \right\rangle = \sum_{[\alpha ,\beta ,\gamma ]\in {\cal
M}}\omega _{\beta ,\gamma } H(\alpha ,\beta ,\gamma ).
\end{eqnarray}
Here  $\langle \, \cdot\, ,
\cdot \, \rangle $ is the Killing form of $\fr{g}$.
In the presence of resonance,  the interaction Hamiltonian \eqref{eq:H-i} is written as a sum of terms corresponding to ``elementary decays'' (see Figure \ref{fig:decay}): to each ``elementary decay'' $q_\alpha \rightarrow q_\beta + q_\gamma$ ($\alpha=\beta+\gamma$ and $\alpha,\beta,\gamma \in \Delta$) there corresponds a term in the interaction Hamiltonian of the form
\begin{equation}\label{eq:decay}
H(\alpha ,\beta ,\gamma ) = \int_{-\infty}^{\infty } dx\,
(q_\alpha s_\beta s_\gamma  - s_\alpha q_\beta q_\gamma ), \quad \omega
_{\beta ,\gamma } ={4N_{\beta ,\gamma }\over (\alpha ,\alpha )}\,
\mbox{det}\left(\begin{array}{cc} ({\bf a},\beta ) & ({\bf a},\gamma )
\\ ({\bf b},\beta ) & ({\bf b},\gamma ) \end{array} \right).
\end{equation}
The coupling constants $\omega _{\beta ,\gamma } $ depend on the vectors ${\bf a}$ and ${\bf b}$, dual to the Cartan elements $J$ and $K$ respectively, as well as on the structural constants $N_{\beta ,\gamma }$ of the Cartan-Weyl basis.

The canonical symplectic form $\Omega ^{(0)}$ for \eqref{eq:N-waves} can be written as
\begin{eqnarray}\label{eq:Ome}
\Omega^{(0)} = \sum_{\alpha \in \Delta _+} i{2({\bf a},\alpha ) \over (\alpha ,\alpha )}
\int_{-\infty }^{\infty }dx\, \delta q_{\alpha }(x,t) \wedge
\delta s_{\alpha }(x,t).
\end{eqnarray}
The direct and inverse scattering transform of \eqref{eq:L} together with appropriately chosen boundary conditions ensure the integrability of \eqref{eq:N-waves} via the ISM.
The simplest and best studied is the class of vanishing boundary conditions (see \cite{ZMNP,CaDe,CaDe3,FaTa,gvy08,G1,G2,GC,Kaup,Symm} and the numerous
references therein). Quite well are studied also class of periodic boundary conditions, especially the
class of finite gap and algebro-geometric solutions \cite{ZMNP,AblSe}.

Our aim here is to outline  and deepen the results on the class of constant boundary
conditions (CBC):
\begin{equation}\label{eq:cbc}\begin{split}
 \lim_{x\to \pm \infty} U(x,t,\lambda) \equiv U_\pm (\lambda) = Q_\pm - \lambda J,
\end{split}\end{equation}
where $Q_\pm$ are constant matrices. Starting with the seminal paper of Zakharov and Shabat \cite{ZaSha*73}, soliton equations with CBC have been extensively studied \cite{EKu,113s,ABP,Dokt,zl131,GeSmi,KonVek1,Ku86,Leon,OW,ABP2,PBT,gg1}. We will consider here the $N$-wave hierarchy related to symplectic Lie algebras $\fr(g)\simeq sp(2r,{\Bbb C})$. Its set of positive roots is
\[ \Delta_+ \equiv \{ e_j - e_k, \quad e_j + e_k, \quad 1 \leq j < k \leq r \} \cup \{ 2e_j, \quad 1\leq j \leq r\}. \]
In addition, we will impose the standard symmetry of the Lax operator $s_\alpha = -q_\alpha^*$, $\alpha \in \Delta_+$. This will ensure that if the potential $U(x,t)={\rm ad}_J\, Q(x,t)
- \lambda J$ is a Hermitian matrix, then the Lax operator \eqref{eq:L} is self adjoint. One of the important facts that ensures the integrability of the relevant NLEE
is, that the continuous spectrum of $L$ coincides with the spectra of the asymptotic operators $L_\pm$.

 Typically the continuous spectrum of these operators
fill up two intervals $(-\infty , -\rho] \cup [\rho, \infty)$ on the real axis in the complex $\lambda$-plane,
while the discrete eigenvalues $\lambda_j$ are in the lacuna $[-\rho, \rho]$. The analyticity properties of the
columns of the Jost solutions are on the Riemannian surfaces $\mathfrak{s}_k$ determined by $\sqrt{a_k^2\lambda^2 - \rho_k^2}$. The best
and most complete description of this issues which are fundamental for the scalar NLS equation with CBC is
contained in the monograph \cite{FaTa}. Of course treating similar problems for the multicomponent NLS (MNLS) is more
difficult, although in many multicomponent cases the analyticity properties are on single $\mathfrak{s}$, \cite{Manak1,ABP,KRB,ABP2,PBT}.

Below we extend these results for systems whose spectral properties are determined by at least two Riemannian surfaces $\mathfrak{s}_1 \cup
\mathfrak{s}_2$. That is why we treat Lax operators whose potentials $Q(x,t)$ belongs to simple Lie algebra $\mathfrak{g}$ with rank 2
and higher.

The present paper continues the work done in \cite{gg1}. The structure of the paper is as follows:   In Section 2  we first consider the spectra of the asymptotic operators for different types of CBC, i.e. for different choices of the asymptotic values of the potential $Q_\pm = \lim_{x\to \pm \infty} Q(x)$. We find that for generic
choice of $Q_\pm$ the analyticity properties are related to Riemannian surface of genus $1$ or higher. We specify
$Q_\pm$ for which the  Riemannian surface is of genus 0 and is union of two or more  $\cup_k \mathfrak{s}_k $. In the same Section
we derive the integral equations for the Jost solutions. Next, following the ideas of Shabat \cite{Sha1, Sha2} we modify
them into integral equations for the fundamental analytic solutions (FAS) of $L$. This generalizes the results of \cite{zl131}
to  simple Lie algebras of the type $sp(2r,{\Bbb C})$.
In Section 4  we generalize the Wronskian relations which allow us  to analyze the mapping
between the scattering data and the $x$-derivative of the potential $Q_x$.
Next, using the Wronskian relations we derive the dispersion laws for the
$N$-wave hierarchy and describe the NLEE related to the given Lax operator.

\section{$N$-wave hierarchy with CBC}\label{sec:2}

We will outline here the direct scattering transform for Lax operators \eqref{eq:L} and will  describe their spectral properties. This will be done on examples of $3$- and $4$-wave systems, related to the algebras  $sl(3,{\Bbb C})$ and $sp(4,{\Bbb C})$, respectively.

\subsection{Soliton equations related to $sl(3)$}

If we take ${\fr g}\simeq sl(3,{\Bbb C})$, we can write potential matrix of the Lax operator \eqref{eq:L} takes the form
\begin{equation}\label{eq:3wa}\begin{split}
Q(x,t) = \left(\begin{array}{ccc} 0 & q_1 & q_3 \\ s_1 & 0 & q_2 \\ s_3 & s_2 & 0  \end{array}\right), \qquad
J = \diag (a_1, a_2, a_3), \quad K = \diag (b_1, b_2, b_3).
\end{split}\end{equation}
We presume that the potential matrices $Q(x,t)$, $J$ and $K$ are traceless (i.e. the Lax operators take values in the algebra $sl(3, {\Bbb C})$) and the eigenvalues of $J$ and $K$ are ordered as follows: $a_1>a_2>a_3$, $b_1>b_2>b_3$ ($a_1+a_2+a_3=0$ and $b_1+b_2+b_3=0$). Here $\lambda \in {\Bbb C}$ is a spectral parameter. This results to the following system of 3 equations (plus another set of ``conjugate equations'', obtained by swapping $q_k$'s and $s_k$'s):
\begin{equation}\label{eq:3wa'}\begin{split}
& i (a_1 - a_2) \frac{\partial q_1}{ \partial t} -  i (b_1 - b_2) \frac{\partial q_1}{ \partial x} - \kappa q_3 s_2 =0, \\
& i (a_2 - a_3) \frac{\partial q_2}{ \partial t} -  i (b_2 - b_3) \frac{\partial q_2}{ \partial x} - \kappa q_3 s_1 =0, \\
& i (a_1 - a_0) \frac{\partial q_3}{ \partial t} -  i (b_1 - b_3) \frac{\partial q_3}{ \partial x} + \kappa q_1 q_2 =0,
\end{split}\end{equation}
where the interaction (coupling) constant reads
\[
\kappa = a_1 (b_2-b_3) +  a_2 (b_3-b_1) +  a_3 (b_1-b_2).
\]
The quantities
\[
v_1 = {b_1-b_2\over a_1-a_2}, \qquad v_2 = {b_2-b_3\over a_2-a_3}, \qquad v_3 = {b_1-b_3\over a_1-a_3}
\]
represent the group velocities of the wave triple $(q_1,q_2,q_3)$ \cite{ZMNP}.

If we impose the symmetry $s_k = -q_k^*$ on the potential \eqref{eq:3wa}, the system \eqref{eq:3wa'} will reduce to the standard $3$-wave interaction model \cite{ZM,vgrn,K,KRB}:
\begin{equation}\label{eq:3waa}\begin{split}
& i (a_1 - a_2) \frac{\partial q_1}{ \partial t} -  i (b_1 - b_2) \frac{\partial q_1}{ \partial x} + \kappa q_3 q_2^* =0, \\
& i (a_2 - a_3) \frac{\partial q_2}{ \partial t} -  i (b_2 - b_3) \frac{\partial q_2}{ \partial x} + \kappa q_3 q_1^* =0, \\
& i (a_1 - a_0) \frac{\partial q_3}{ \partial t} -  i (b_1 - b_3) \frac{\partial q_3}{ \partial x} + \kappa q_1 q_2 =0,
\end{split}\end{equation}
The (canonical) Hamiltonian  of (\ref{eq:3wa'}) is given by \cite{ZM,vgrn,ZMNP}:
\begin{eqnarray}\label{eq:1.5-3w}
H_{\rm 3-w} ={1 \over 2}\int_{-\infty }^{\infty }  dx\, \left(  \sum_{k=1}^{3} v_k\left(  q_k \frac{\partial q_k^*}{ \partial x }
- q_k^* \frac{\partial q_k}{ \partial x }\right) +\kappa (q_3 q_1^* q_2^* + q_3^*q_1q_2 ) \right).
\end{eqnarray}
The decay diagram corresponding to this process is given on Figure \ref{fig:decay}.

In the case of CBC, there correspond the so-called ``dark solitons'' whose
properties and behavior substantially differ from the ones of the ``bright
solitons'' (corresponding to zero boundary conditions) \cite{FaTa}. The dark solitons for the nonlinear Schr\"odinger type equations and their generalizations  with non-vanishing boundary conditions are studied in \cite{gk78,zl131,Leon,KonVek1,ABP2,PBT}. Similar results for the discrete nonlinear Schr\"odinger type equations (the Ablowitz-Ladik hierarchy) are obtained in \cite{ABP}.

\subsection{Soliton equations related to $sp(4)$}

The set of positive roots of $sp(4,{\Bbb C})$ are given by  \cite{Helg}: $\Delta_+ =\{ e_1 - e_2, e_1, e_2, e_1 + e_2\} $.
We will use the typical representation of $sp(4,{\Bbb C})$. In this representation the potential matrix of the Lax operator \eqref{eq:L} takes the form
\begin{equation}\label{eq:sp4u} \begin{split}
 Q(x,t) = \left(\begin{array}{cccc} 0 & q_4 & q_3 & q_1 \\ s_4 & 0 & q_2 & q_3 \\ s_3 & s_2 & 0 & q_4 \\
 s_1 & -s_3 & s_4 &  0   \end{array}\right), \qquad
\begin{array}{c}   J = \diag (a_1, a_2, -a_2, -a_1), \\ \\   K = \diag (b_1, b_2, -b_2, -b_1).  \end{array}
\end{split}\end{equation}
Imposing the symmetry $s_k = -q_k^*$ on the potential \eqref{eq:sp4u} will result to the $4$-wave system:
\begin{equation}\label{eq:4-wave}\begin{split}
& i (a_1 - a_2) \frac{\partial q_4}{ \partial t} -  i (b_1 - b_2) \frac{\partial q_4}{ \partial x} +2 \kappa (q_1 q_3^* - q_2^* q_3) =0, \\
& i (a_1 + a_2) \frac{\partial q_3}{ \partial t} -  i (b_1 + b_2) \frac{\partial q_3}{ \partial x} +2 \kappa (-q_1 q_4^* + q_4 q_2) =0, \\
& i a_1  \frac{\partial q_1}{ \partial t} -  i b_1  \frac{\partial q_1}{ \partial x} - 2 \kappa q_3 q_4 =0, \\
& i a_2  \frac{\partial q_2}{ \partial t} -  i b_2  \frac{\partial q_2}{ \partial x} - 2 \kappa q_4^*q_3 =0,
\end{split}\end{equation}
with a coupling constant $\kappa =a_1b_2-a_2b_1 $. The system \eqref{eq:4-wave} is described by the interaction
 (canonical) Hamiltonian:
\begin{eqnarray}\label{eq:H4}
H_{\rm 4w,int} = 2\kappa\int_{-\infty }^{\infty } dx \,\left[
 \left( q_{3}q_4^*q_{2}^* + q_{3}^*q_4q_{2} \right)
-  \left( q_{1}q_{2}^*q_{3}^* +  q_{1}^*q_{2}q_{3}\right)
\right].
\end{eqnarray}
If we identify  $q_{4} = Q$, $q_{3} =
E_p$, $q_{1} = E_a$ and $q_{2} =E_s$, where $Q $ is the normalized
effective polarization of the medium and $E_p $, $E_s $ and $E_a $ are the
normalized pump, Stokes and anti-Stokes wave amplitudes respectively,
then we obtain the system of equations generalizing the one studied
in \cite{3} which describes Stokes--anti-Stokes stimulated Raman scattering.

\begin{figure}
\begin{center}
\begin{tikzpicture}
\shade[ball color=blue!20] (0,2) circle (.35cm);
\shade[ball color=blue!20] (-1,0) circle (.35cm);
\shade[ball color=blue!20] (-2,-2) circle (.35cm);
\shade[ball color=blue!20] (2,-2) circle (.35cm);
\draw (0,2) node {$q_1$};
\draw (-1,0) node {$q_3$};
\draw (-2,-2) node {$q_4$};
\draw (2,-2) node {$q_2$};
\draw[thick,-latex] (0,1.6)--(-0.85,0.35);
\draw[thick,-latex] (0,1.6)--(1.85,-1.65);

\draw[thick,-latex] (-1.1,-0.4)--(1.65,-1.9);
\draw[thick,-latex] (-1.1,-0.4)--(-1.95,-1.65);
\end{tikzpicture}
\end{center}
  \caption{\small The  wave-decay diagram corresponding to the cascade interaction Hamiltonian  \eqref{eq:H4}.}\label{fig:decay2}
\end{figure}
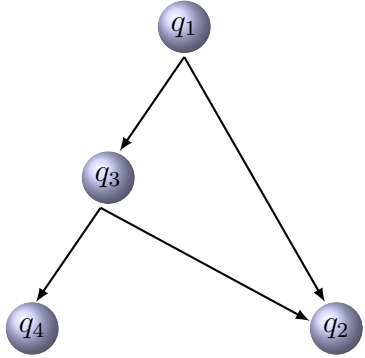

In the case of CBC, the first important constraint which ensures that the problem is solvable consists in the requirement that the asymptotic
operators:
\begin{equation}\label{eq:Lpm}\begin{split}
 L_\pm &\equiv i \frac{\partial \psi_\pm }{ \partial x } + U_\pm (\lambda) \psi_\pm =0, \qquad U_\pm (\lambda) = Q_\pm - \lambda J
 \end{split}\end{equation}
have the same spectra. Here $Q_\pm = \lim_{x\to \pm \infty} Q(x,t)$. In particular this condition means that $U_\pm (\lambda)$
must have the same eigenvalues, i.e. have the same characteristic polynomials.

The next important step is to choose the asymptotic values of $Q_\pm = \lim_{x\to \pm \infty} Q(x,t)$ so that
the above constraint is satisfied. Without being exhaustive we will consider a few examples.

We start with the generic case, when
\begin{equation}\label{eq:U1}\begin{split}
 U_{1,\pm} = \sum_{\alpha \in \Delta_+}^{}( q_{\alpha,\pm} E_\alpha +s_{\alpha,\pm} E_{-\alpha})-\lambda J.
\end{split}\end{equation}
The spectrum of the asymptotic operators $L_\pm$ is determined by the eigenvalues of the corresponding $U_{1,\pm}(\lambda)$.
So we need the characteristic polynomial $C_1(z,\lambda)= \det( z\openone -U_{1,\pm}(\lambda))$ of $U_{1,\pm}(\lambda)$ which comes out to be rather
complicated. One can relate to each  such polynomial an algebraic curve, which in this case has genus 3.

Next we consider less complicated case with
\begin{equation}\label{eq:U2}\begin{split}
 U_{2,\pm} = \left(\begin{array}{cccc} -\lambda a_1 & q_{4,\pm} &  q_{3,\pm} & 0 \\  s_{4,\pm} & - \lambda a_2 & 0 & - q_{3,\pm} \\
  s_{3,\pm} & 0 & \lambda a_2 &  q_{4,\pm} \\ 0 & - s_{3,\pm} &  s_{4,\pm} & \lambda a_1   \end{array}\right).
\end{split}\end{equation}
Now the characteristic polynomial is much simpler:
\begin{equation}\label{eq:C2}\begin{split}
 C_2(z,\lambda) = z^4 - ((a_1^2 +a_2^2) \lambda^2 - 2 (\rho_3^2 + \rho_4^2 )) z^2 +  a_1^2 a_2^2 \lambda^4 - 2 a_1 a_2 ( \rho_3^2 - \rho_4^2 )
 \lambda^2 + 2 \rho_3^2  \rho_4^2 .
\end{split}\end{equation}
Here and below $\rho_k^2  = -|q_{k,\pm}|^2 $.
Now the corresponding curve has genus 1, i.e. the eigenvalues of $L_{2,\pm}$ will be expressed in terms of Jacobi elliptic functions.

The third case that we will consider is
\begin{equation}\label{eq:U}\begin{split}
 U_{\pm} = \left(\begin{array}{cccc} -\lambda a_1 & 0 &  0 & q_{1,\pm}  \\  0 & - \lambda a_2 & q_{2,\pm} & 0 \\
 0 &  s_{2,\pm} & \lambda a_2 &  0 \\   s_{1,\pm} & 0 &  0 & \lambda a_1   \end{array}\right).
\end{split}\end{equation}
The characteristic polynomial now can be factorized into:
\begin{equation}\label{eq:Czla}\begin{split}
 C(z,\lambda) = (z^2 - \lambda^2 a_1^2 + \rho_1^2) (z^2 - \lambda^2 a_2^2 + \rho_2^2).
\end{split}\end{equation}
The corresponding curve now is a direct product of two genus 0 curves. This is the simplest nontrivial case which we
will focus on below. Each of this curves is related to a Riemannian surface determined by $\sqrt{a_1^2 \lambda^2 -\rho_1^2}$
and $\sqrt{a_2^2 \lambda^2 -\rho_2^2}$.

Indeed, the eigenvalues of the asymptotic operators are $\pm z_k(\lambda)$ where
\begin{equation}\label{eq:zla}\begin{split}
 z_{k}(\lambda) =  \sqrt{a_k^2\lambda^2-\rho_k^2}.
\end{split}\end{equation}
The corresponding eigenvectors of $U_\pm$ can be combined into the matrix
\begin{equation}\label{eq:fi0}\begin{split}
 \phi_{0,\pm} &= \left(\begin{array}{cccc} -q_{1,\pm} & 0 & 0 & \frac{1}{2z_1} \\ 0 & - q_{2,\pm } &  \frac{1}{2z_2} & 0 \\
 0 & -\lambda a_2 +z_2 & \frac{\lambda a_2 +z_2  }{2q_{2,\pm} z_2} & 0 \\ -\lambda a_1 +z_1 & 0 & 0 & \frac{\lambda a_1 +z_1 }{2q_{1,\pm}z_1}
  \end{array}\right), \\
 \phi^{-1}_{0,\pm} &= \left(\begin{array}{cccc} -\frac{\lambda a_1 +z_1 }{2q_{1,\pm}z_1} & 0 & 0 & \frac{1}{2z_1} \\
 0 & -\frac{\lambda a_2 +z_2 }{2q_{2,\pm}z_2} &  \frac{1}{2z_2} & 0 \\
 0 & -\lambda a_2 +z_2 & q_{2,\pm}  & 0 \\ -\lambda a_1 +z_1 & 0 & 0 & q_{1,\pm}   \end{array}\right).
\end{split}\end{equation}
Note that $\phi_{0,\pm} \in SP(4)$ and must fulfill
\begin{equation}\label{eq:Upm0}\begin{split}
U_\pm (\lambda) = - \phi_{0,\pm}^{-1} \mathcal{J}(\lambda)\phi_{0,\pm}, \qquad
\mathcal{J} =  z_1 H_{1}+ z_2 H_{2} = \diag ( z_1, z_2, -z_2, -z_1).
\end{split}\end{equation}

\begin{figure}

\begin{center}

\begin{tikzpicture}[scale=1.25]


\draw[->] (-3,0) -- (3,0);

\draw[->] (0,-2.5) -- (0,2.5);

\draw[magenta, ultra thick] (-2.75,0) -- (-1.75,0);

\draw[magenta, ultra thick] (1.75,0) -- (2.75,0);

\draw[magenta,  thick] (-2.75,0) -- (-1,0);

\draw[magenta,  thick] (1,0) -- (2.75,0);

\fill[magenta] (-1.75,0) circle (2pt);

\fill[magenta] (1.75,0) circle (2pt);

\fill[magenta] (-1,0) circle (1.5pt);

\fill[magenta] (1,0) circle (1.5pt);


\draw (-0.25,-0.25) node {\small $O$};

\draw (-1.8,-0.25) node {\small $-\rho_1$};

\draw (1.8,-0.25) node {\small $\rho_1$};

\draw (-1,-0.25) node {\small $-\rho_2$};

\draw (1,-0.25) node {\small $\rho_2$};


\draw[magenta] (2.25,2.25) node {\small $\lambda$};

\draw(0,-3) node {\small {\bf (a)}};
\end{tikzpicture} \quad
\begin{tikzpicture}[scale=1.25]


\draw[->] (-3,0) -- (3,0);

\draw[->] (0,-2.5) -- (0,2.5);


\draw[blue, thick,] (2,0) arc (0:360:2);

\draw[blue, ultra  thick,] (2,0) arc (0:30:2);

\draw[blue, ultra  thick,] (2,0) arc (0:-30:2);

\draw[blue, ultra thick,rotate=180] (2,0) arc (0:30:2);

\draw[blue, ultra  thick,rotate=180] (2,0) arc (0:-30:2);

\draw[blue ] (-2.75,0) -- (2.75,0); 

\fill[blue] (-2,0) circle (2pt);

\fill[blue] (2,0) circle (2pt);
\fill[blue] (-1.732,1) circle (2pt);

\fill[blue] (-1.732,-1) circle (2pt);

\fill[blue] (1.732,1) circle (2pt);

\fill[blue] (1.732,-1) circle (2pt);


\draw (-0.25,-0.25) node {\small $O$};

\draw (-2.4,-0.25) node {\small $-A$};

\draw (2.25,-0.25) node {\small $A$};

\draw (2.2,1) node {\small $B^+$};

\draw (2.2,-1) node {\small $\bar{B}^+$};

\draw (-2.2,1) node {\small $B^-$};

\draw (-2.2,-1) node {\small $\bar{B}^-$};


\draw[blue] (2.25,2.25) node {\small $z_1(\lambda)$};

\draw(0,-3) node {\small {\bf (b)}};
\end{tikzpicture}

\end{center}

  \caption{ The spectrum of the asymptotic operators $L_\pm$. On the left panel {\bf (a)} it is located on the two pairs of cuts
on the complex $\lambda$-plane,  that determine the Riemannian surfaces $s_1\cup s_2$; on the right panel {\bf (b)} we show it
on the complex $z_1$-plane, where $z_1$ is the uniformizing variable for the first Riemannian surface $s_1$; see also Remark \ref{rem:1} below.}\label{fig:6s-5}

\end{figure}
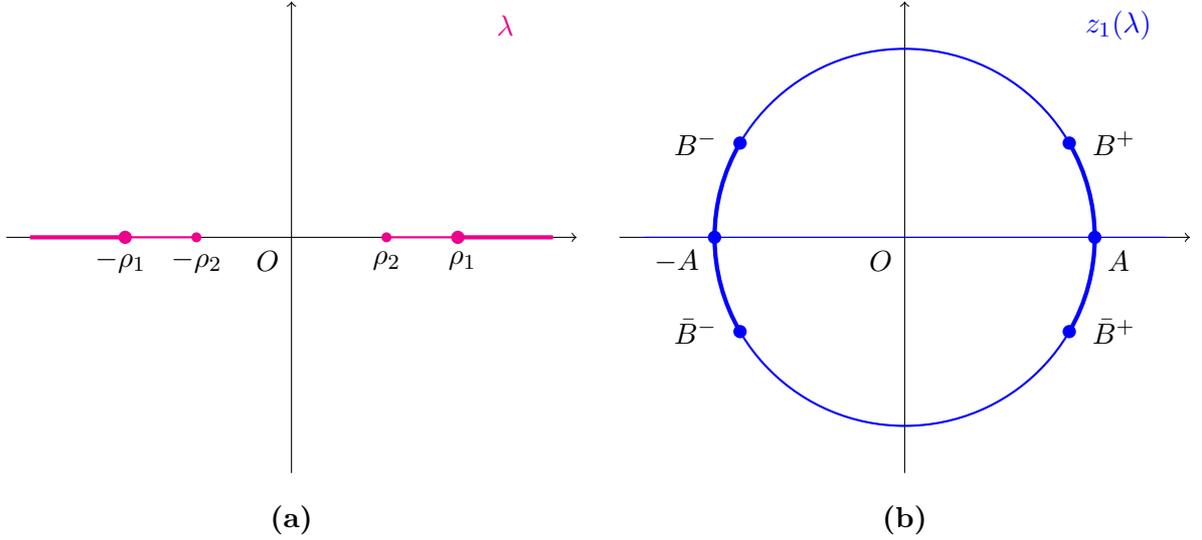

\begin{remark}\label{rem:1}
 The continuous spectrum of the Lax operator coincides with the spectrum of the asymptotic operators. It is
 shown by thick lines on Figure \ref{fig:6s-5}. The first special feature of the spectrum is that it has varying
 multiplicity. Indeed, on the cuts $(-\infty, -\rho_1] \cup [\rho_1, \infty)$ the multiplicity is 4;
 on the cuts $(-\rho_1, -\rho_2] \cup [\rho_2, \rho_1)$ the multiplicity is 2; on the lacuna the multiplicity is 0
 (panel (a)). The same feature is present also on panel (b): the spectrum has multiplicity 4 on the cuts
 $(-\infty, -\rho_1] \cup [\rho_1, \infty)$ and multiplicity 2 on the arcs $(-\bar{B}^-, B^-) \cup (\bar{B}^+ , B^+)$.
 We may have two types of discrete eigenvalues: on panel (a) type 1 eigenvalues are located in  $(-\rho_1, -\rho_2] \cup [\rho_2, \rho_1)$ and orthogonal
 to the continuous spectrum in that intervals;  type 2 are located in the lacuna $[-\rho_2, \rho_2]$. On panel (b) these
 eigenvalues are located on the arcs $(-\bar{B}^-, B^-) \cup (\bar{B}^+ , B^+)$, while type 2 eigenvalues are located
 on the arcs $(\bar{B}^-, B^+) \cup (\bar{B}^- , \bar{B}^+)$. In addition we may have virtual solitons located at the end points of the spectrum.
 On panel (a) these are $\pm \rho_1$ and $\pm \rho_2$; on panel (b) they are at the points- $\pm A$ and $B^\pm, \bar{B}^\pm$.

\end{remark}

\subsection{Jost solutions and FAS of $L$}

The starting point in developing the direct scattering transform for \eqref{eq:N-waves} are the eigenfunctions (the so-called Jost solutions) of L \eqref{eq:L},
determined uniquely by their  asymptotics for $x\to \pm \infty$:
\begin{equation}\label{eq:Jost0}\begin{split}
\lim _{x\to \infty} \psi(x,t,\lambda) e^{i \mathcal{J}(\lambda) x} \hat{\phi}_{0,+} = \openone , \qquad
\lim _{x\to -\infty} \phi(x,t,\lambda) e^{i \mathcal{J}(\lambda) x} \hat{\phi}_{0,-} = \openone .
\end{split}\end{equation}
The scattering matrix $T(t,\lambda) $ is introduced by:
\begin{equation}\label{eq:T}\begin{split}
 \phi(x,t,\lambda) = \psi(x,t,\lambda) T(t,\lambda).
\end{split}\end{equation}
If we now introduce:
\begin{equation}\label{eq:Xpm}\begin{split}
 X_+(x,t,\lambda) = \psi(x,t,\lambda) e^{i \mathcal{J}(\lambda) x} \hat{\phi}_{0,+} , \qquad
 X_-(x,t,\lambda) = \phi(x,t,\lambda) e^{i \mathcal{J}(\lambda) x} \hat{\phi}_{0,-},
\end{split}\end{equation}
then it is not difficult to check that $X_\pm$ satisfy the equation:
\begin{equation}\label{eq:Xpm1}\begin{split}
 i \frac{\partial X_\pm}{ \partial x } + (Q(x,t) - Q_\pm) X_\pm (x,t,\lambda) +
 \left[ Q_\pm - \lambda J, X_\pm (x,t,\lambda) \right] =0.
\end{split}\end{equation}
Multiplying eq. (\ref{eq:Xpm1}) by $\hat{\phi}_{0,\pm}$ on the left and by $\phi_{0,\pm}$ on the
right we obtain:
\begin{equation}\label{eq:Xpm2}\begin{split}
 i \frac{\partial \tilde{X}_\pm}{ \partial x } + \tilde{U}_\pm(x,t,\lambda)  \tilde{X}_\pm (x,t,\lambda) -
 \left[ \mathcal{J}(\lambda), \tilde{X}_\pm (x,t,\lambda) \right] =0,
\end{split}\end{equation}
where $\tilde{X}_\pm \equiv \hat{\phi}_{0,\pm} X_\pm \phi_{0,\pm}$ and
\begin{equation}\label{eq:tilU}\begin{split}
\tilde{U}_\pm(x,t,\lambda) = \hat{\phi}_{0,\pm}(Q(x,t) - Q_\pm) \phi_{0,\pm} .
\end{split}\end{equation}
From eqs. (\ref{eq:Xpm2}) one can derive the following integral equations for the Jost solutions:
\begin{equation}\label{eq:Jost1}\begin{split}
 \tilde{X}_\pm = \openone + i \int_{\pm \infty}^{x} dy\; e^{-i \mathcal{J}(\lambda) (x-y) } \tilde{U}_\pm (y,t,\lambda)
 \tilde{X}_\pm (y,t,\lambda)  e^{i \mathcal{J}(\lambda) (x-y) } ,
\end{split}\end{equation}
or in components we have:
\begin{equation}\label{eq:Jost2}\begin{split}
 (\tilde{X}_\pm)_{jk} = \delta_{jk} + i \int_{\pm \infty}^{x} dy\; e^{-i (\mathcal{J}_j -\mathcal{J}_k)  (x-y) }
 \left( \tilde{U}_\pm (y,t,\lambda) \tilde{X}_\pm (y,t,\lambda)\right)_{jk}.
\end{split}\end{equation}
It is natural to assume that the potential $Q(x,t)$ is smooth enough and that  $Q(x,t)- Q_\pm$ tend to 0
for $x\to \pm \infty$ faster than any polynomial of $x$. Therefore eqs. (\ref{eq:Jost2}) have solutions for all
$\lambda$, for which $\im (\mathcal{J}_j -\mathcal{J}_k)=0 $. Then the exponential factors in (\ref{eq:Jost1}) are
bounded and the integrals in the right hand side of (\ref{eq:Jost1}) converge due to the above condition on $Q(x,t)$.
Thus the Jost solutions of $L$ are well defined on the lines in the complex $\lambda$-plane for
which  $\im (\mathcal{J}_j -\mathcal{J}_k)=0 $, which constitute the continuous spectrum of $L$.

Let us now outline the possibility to extend analytically in $\lambda$ some of the columns of the Jost solutions.
Such extension  for the $j$-th column of $\tilde{X}_\pm$ in the region $\lambda \in \mathcal{A}_\pm$ will be possible if one can ensure that all exponential factors
$\mathcal{E}_{jk}(\lambda) = \exp( -i( \mathcal{J}_j - \mathcal{J}_k)(x-y))$ fall off for $x- y \to \pm \infty$ and $\lambda \in \mathcal{A}_\pm$. We will define $\mathcal{A}_\pm$ as follows:
\begin{equation}\label{eq:spm}\begin{split}
 \mathcal{A}_+ \equiv \{ \im (\mathcal{J}_j -\mathcal{J}_k) > 0, \quad j<k \}, \qquad  \mathcal{A}_- \equiv \{ \im (\mathcal{J}_j -\mathcal{J}_k) < 0, \quad j<k\}.
\end{split}\end{equation}
It is easy to check that the first columns of $\psi(x,t,\lambda)$ and  $\phi(x,t,\lambda)$ allow analytic extensions for $\lambda \in \mathcal{A}_-$ and
 $\lambda \in \mathcal{A}_+$ respectively; analogously, the last columns of $\psi(x,t,\lambda)$ and  $\phi(x,t,\lambda)$ allow analytic extensions for $\lambda \in \mathcal{A}_+$ and $\lambda \in \mathcal{A}_-$ respectively. As to the rest of the columns of the Jost solutions we find, that some of their exponential factors
 fall off, while the others blow up for $x-y \to \pm \infty$. Thus they do not allow analytic extensions in $\lambda$.

 The construction of FAS however is possible using Shabat's method \cite{PBT,Sha1}, see also \cite{ZMNP,gg1}. It consists in modifying the integral equations (\ref{eq:Jost2}) into:
\begin{equation}\label{eq:FAS1}\begin{aligned}
(\tilde{Y}_+)_{jk} &=  i \int_{ \infty}^{x} dy\; e^{-i (\mathcal{J}_j -\mathcal{J}_k)  (x-y) }
 \left( \tilde{U}_+ (y,t,\lambda) \tilde{Y}_+ (y,t,\lambda)\right)_{jk}  , &\quad j &<k;\\
 (\tilde{Y}_+)_{jk} &= \delta_{jk} + i \int_{- \infty}^{x} dy\; e^{-i (\mathcal{J}_j -\mathcal{J}_k)  (x-y) }
 \left( \tilde{U}_- (y,t,\lambda) \tilde{Y}_+ (y,t,\lambda)\right)_{jk}  ,&\quad j & \geq k;
\end{aligned}\end{equation}
and
\begin{equation}\label{eq:FAS2}\begin{aligned}
(\tilde{Y}_-)_{jk} &= \delta_{jk} + i \int_{ -\infty}^{x} dy\; e^{-i (\mathcal{J}_j -\mathcal{J}_k)  (x-y) }
 \left( \tilde{U}_- (y,t,\lambda) \tilde{Y}_- (y,t,\lambda)\right)_{jk}  , &\quad j & \geq k;\\
 (\tilde{Y}_-)_{jk} &=  i \int_{ \infty}^{x} dy\; e^{-i (\mathcal{J}_j -\mathcal{J}_k)  (x-y) }
 \left( \tilde{U}_+ (y,t,\lambda) \tilde{Y}_- (y,t,\lambda)\right)_{jk}  ,&\quad j & < k.
\end{aligned}\end{equation}
The integral equations (\ref{eq:FAS1}) and (\ref{eq:FAS2}) give rise to two FAS $\chi^+(x,t,\lambda)$ and $\chi^-(x,t,\lambda)$.
It is well known that any two fundamental solutions of the same operator $L$ are linearly related. That means that we can
find the relations between $\chi^\pm(x,t,\lambda)$ and the Jost solutions of $L$. This can be done by comparing their asymptotics
for $x\to \pm \infty$ with the asymptotics of the Jost solutions. Skipping the details we obtain:
\begin{equation}\label{eq:chipm}\begin{split}
\chi^+(x,t,\lambda) = \phi (x,t,\lambda) S^+(t,\lambda) = \psi (x,t,\lambda) T^-(t,\lambda) D^+(\lambda), \\
\chi^-(x,t,\lambda) = \phi (x,t,\lambda) S^-(t,\lambda) = \psi (x,t,\lambda) T^+(t,\lambda) D^-(\lambda),
\end{split}\end{equation}
where the factors $S^\pm (t,\lambda)$, $T^\pm (t,\lambda)$ and $D^\pm (\lambda)$ are related to the scattering matrix $T (t,\lambda)$ by:
\begin{equation}\label{eq:gaus}\begin{split}
T (t,\lambda) = T^- (t,\lambda) D^+ (\lambda) \hat{S}^+ (t,\lambda) = T^+ (t,\lambda) D^- (\lambda) \hat{S}^- (t,\lambda).
\end{split}\end{equation}
Let us note that from the integral equations (\ref{eq:FAS1}) and (\ref{eq:FAS2}) it follows that the factors $S^+(t,\lambda)$ and
$T^+(t,\lambda)$ are upper-triangular with units on the diagonal; likewise, $S^-(t,\lambda)$ and $T^-(t,\lambda)$ are lower-triangular
with units on the diagonal. The matrices $D^\pm (\lambda)$ are diagonal. As we will find below, $D^\pm (\lambda)$ are time independent;
they are generating functions of the integrals of motion of the corresponding NLEE. Keeping in mind that the Jost solutions, as well
as the FAS of $L$ must belong to the group $SP(2r)$ we find that the factors $S^\pm (t,\lambda)$, $T^\pm (t,\lambda)$ and $D^\pm (\lambda)$
must have the form:
\begin{equation}\label{eq:gaus0}\begin{aligned}
S^\pm (t,\lambda) &= \exp \left( \sum_{\alpha\in \Delta_+}^{} \boldsymbol{\tau}^\pm_\alpha E_{\pm \alpha} \right), &\quad
T^\pm (t,\lambda) &= \exp \left( \sum_{\alpha\in \Delta_+}^{} \boldsymbol{\rho}^\mp_\alpha E_{\pm \alpha} \right),\\
D^\pm (\lambda) & = \exp \left( \sum_{j=1}^{r} d_j^\pm (\lambda) H_j \right).
\end{aligned}\end{equation}
The problem of splitting the scattering matrix into triangular and diagonal factors (\ref{eq:gaus}) is known also as Gauss decomposition.
It has unique solution for all simple Lie algebras, see e.g. the appendix of \cite{Contemp}. As a results all the coefficients $\tau^\pm (t,\lambda)$,
$\rho^\pm (t,\lambda)$ and $d_j^\pm (\lambda)$ cab be expressed explicitly through the matrix elements of the scattering matrix $T(t,\lambda)$.

We can derive two more FAS slightly modifying the integral equations (\ref{eq:FAS1}) and (\ref{eq:FAS2})
\begin{equation}\label{eq:FAS3}\begin{aligned}
(\tilde{Y}'_+)_{jk} &=  i \int_{ \infty}^{x} dy\; e^{-i (\mathcal{J}_j -\mathcal{J}_k)  (x-y) }
 \left( \tilde{U}_+ (y,t,\lambda) \tilde{Y}'_+ (y,t,\lambda)\right)_{jk}  , &\quad j & \leq k;\\
 (\tilde{Y}_+')_{jk} &= \delta_{jk} + i \int_{- \infty}^{x} dy\; e^{-i (\mathcal{J}_j -\mathcal{J}_k)  (x-y) }
 \left( \tilde{U}_- (y,t,\lambda) \tilde{Y}'_+ (y,t,\lambda)\right)_{jk}  ,&\quad j & > k;
\end{aligned}\end{equation}
and
\begin{equation}\label{eq:FAS4}\begin{aligned}
(\tilde{Y}'_-)_{jk} &= \delta_{jk} + i \int_{ -\infty}^{x} dy\; e^{-i (\mathcal{J}_j -\mathcal{J}_k)  (x-y) }
 \left( \tilde{U}_- (y,t,\lambda) \tilde{Y}'_- (y,t,\lambda)\right)_{jk}  , &\quad j & > k;\\
 (\tilde{Y}'_-)_{jk} &=  i \int_{ \infty}^{x} dy\; e^{-i (\mathcal{J}_j -\mathcal{J}_k)  (x-y) }
 \left( \tilde{U}_+ (y,t,\lambda) \tilde{Y}'_- (y,t,\lambda)\right)_{jk}  ,&\quad j & \leq k.
\end{aligned}\end{equation}
The relations of these FAS to the Jost solutions are slightly different from (\ref{eq:chipm})
\begin{equation}\label{eq:chipmp}\begin{split}
\chi^{+,\prime}(x,t,\lambda) = \phi (x,t,\lambda) S^+(t,\lambda)\hat{D}^+(\lambda) = \psi (x,t,\lambda) T^-(t,\lambda), \\
\chi^{+,\prime}(x,t,\lambda) = \phi (x,t,\lambda) S^-(t,\lambda)\hat{D}^-(\lambda) = \psi (x,t,\lambda) T^+(t,\lambda).
\end{split}\end{equation}
Note that the diagonal factors $D^\pm (\lambda)$ are analytic functions of $\lambda$ for $\lambda \in \mathcal{A}_\pm$.

\section{Wronskian relations for Lax operators  with CBC}\label{sec:3}

There is substantial difference between the  NLEEs with vanishing boundary conditions  and the NLEE with CBC.
The potentials $Q(x,t)$ of the Lax operators in both cases may be considered as local coordinates  of the relevant phase space.
For vanishing boundary conditions the corresponding phase space $\Phi_{\rm VBC}$ is linear, while for the CBC case the space
$\Phi_{\rm CBC}$ is {\it nonlinear}. As we shall see below, this fact substantially changes the dispersion law of the
NLEE.

Another problem that arises from it, is the interpretation of the ISM as a generalised Fourier transform. This idea was proposed in the seminal AKNS paper
\cite{AKNS}; in the next two decades it was shown to be valid to a large class of Lax operators, see \cite{Contemp, G*86, vgn2}. In the  case of vanishing boundary conditions,
one was able to derive expansions for the potential $Q(x,t)$ over the ``squared solutions'' of $L$. Of course, such a generalised Fourier transform hold true only
for linear spaces. In fact, the completeness of the `squared solutions` was derived also for the Zakharov-Shabat system with CBC
\cite{KonVek1}, but no expansion of $Q(x,t)$ and description of the class  of NLEE s could be presented.

Our idea is to use generalized Wronskian relations, which will allow us to study the mapping between $x$-derivative of the potential
$Q_x(x,t)$ and the scattering data. Note, that $Q_x(x,t)$ is an element from a linear space. Therefore its expansion over the `squared
solutions` of $L$ is legitimate and can be used  for the description of the $N$-wave hierarchy.

\subsection{Wronskian relations for $Q_x$}

The importance of the Wronskian relations for the analysis of the mappings $\mathcal{F}: Q(x,t) \to \mathcal{T}$ between
the potential of $L$ and the scattering data has been noted long ago by Calogero and Degasperis \cite{CaDe1,CaDe2,Ca*89,Ca*89a}.
They were based on the basic relations
\begin{equation}\label{eq:CaDe}\begin{split}
\left.  i \left( \hat{\psi} J \psi \right) \right|_{x=-\infty}^\infty &=
-\int_{-\infty}^{\infty} \hat{\psi} [J, Q(x,t)] \psi \; dx, \\
\left.  i \left( \hat{\psi}  J \delta\psi \right) \right|_{x=-\infty}^\infty &=
-\int_{-\infty}^{\infty} \hat{\psi} \delta Q(x,t) \psi \; dx.
\end{split}\end{equation}
Calogero and Degasperis \cite{CaDe1,CaDe2} also generalized these relations which allowed them to describe
the classes of B\"acklund transformations for the relevant NLEE.

However their approach can not be applied directly for the equations with CBC.
The difficulties come from the fact that the asymptotics of the Jost solutions (\ref{eq:Jost0})
are more complicated than the ones for VBC. Here we propose generalizations of
Calogero and Degasperis approach which allows one to treat also CBC.
One of the important facts is that we are using the FAS instead of the Jost solutions.
The second issue is that we use a modification of (\ref{eq:CaDe}), namely:
\begin{equation}\label{eq:11.1}\begin{split}
\left.  i \left( \hat{\chi} (Q - \lambda J) \chi \right) \right|_{x=-\infty}^\infty &=
\int_{-\infty}^{\infty} \hat{\chi} iQ_x \chi \; dx, \\
\tr \left.  i \left( \hat{\chi} (Q - \lambda J) \chi \right) E_\alpha \right|_{x=-\infty}^\infty &=
\int_{-\infty}^{\infty} \tr  iQ_x \chi E_\alpha \hat{\chi} \; dx =- i\biglb \ad_J^{-1} Q_x , e_\alpha (x,\lambda) \bigrb,
\end{split}\end{equation}
where
\begin{equation}\label{eq:11.3}\begin{split}
e_\alpha^\pm & = \pi_j \chi^\pm E_\alpha \hat{\chi}^\pm (x,\lambda), \qquad \pi_j = \ad_J^{-1} \ad_J, \\
 \biglb X, Y \bigrb &= \int_{-\infty}^{\infty} \tr  (X, [J, Y]) .
\end{split}\end{equation}
The limits in the l.h.side must be expressed by the scattering data, {\it and now the limits $Q_\pm$ do not come up}.
Indeed, the l.h.side of (\ref{eq:11.1}) is expressed using (\ref{eq:chipm}) as follows:
\begin{equation}\label{eq:8.12}\begin{split}
& \lim_{x\to \infty} i \left( \hat{\chi} (Q - \lambda J) \chi \right)= -i \hat{D}^\pm  \hat{T}^\mp \mathcal{J}(\lambda)  T^\mp (t,\lambda) D^\pm(\lambda),
\end{split}\end{equation}
and
\begin{equation}\label{eq:8.13}\begin{split}
& \lim_{x\to -\infty} i \left( \hat{\chi} (Q - \lambda J) \chi \right) = -i \hat{S}^\mp  \mathcal{J}(\lambda)  S^\pm (t,\lambda).
\end{split}\end{equation}
In a similar manner, for the second pair of FAS we have:
\begin{equation}\label{eq:8.12'}\begin{split}
& \lim_{x\to \infty} i \left( \hat{\chi'} (Q - \lambda J) \chi' \right)= -i \hat{T}^\mp  \mathcal{J}(\lambda)  T^\mp (t,\lambda) ,
\end{split}\end{equation}
and
\begin{equation}\label{eq:8.13'}\begin{split}
& \lim_{x\to -\infty} i \left( \hat{\chi} (Q - \lambda J) \chi \right)= -i \hat{D}^\pm  \hat{S}^\pm  \mathcal{J}(\lambda)  S^\pm (t,\lambda)D^\pm.
\end{split}\end{equation}
Let us now multiply both sides of (\ref{eq:11.1}) by $E_\alpha$ and take a Killing form with $E_{\mp\alpha}$.
The left hand sides becomes:
\begin{equation}\label{eq:9.6}\begin{split}
\left.  i\left\langle  \left( \hat{\chi}^\pm (Q - \lambda J) \chi^\pm, E_{\mp\beta} \right\rangle \right) \right|_{x=-\infty}^\infty
&=i \left\langle \hat{S}^\pm \mathcal{J}(\lambda))S^\pm (\lambda) E_{\mp \beta} \right\rangle \\
&= i  \tau_\beta^\pm (\lambda,t) ,\\
\left.  i\left\langle  \left( \hat{\chi}^\pm (Q - \lambda J) \chi^\pm, E_{\pm\beta} \right\rangle \right) \right|_{x=-\infty}^\infty
&=-i \left\langle \hat{D}^\pm \hat{T}^\mp \mathcal{J}(\lambda))T^\mp (\lambda)D^\pm E_{\pm \beta} \right\rangle \\
&= -i  \rho_\beta^\pm (\lambda,t)  \exp(\pm\vec{d}^\pm (\lambda),\beta).
\end{split}\end{equation}
Similarly, using $\chi^{\pm,\prime}$ we have:
\begin{equation}\label{eq:9.4}\begin{split}
\left.  i\left\langle  \left( \hat{\chi}^{\pm,\prime} (Q - \lambda J) \chi^{\pm,\prime}, E_{\mp\beta} \right\rangle \right) \right|_{x=-\infty}^\infty
&=i \left\langle D^\pm \hat{S}^\pm \mathcal{J}(\lambda))S^\pm (\lambda) \hat{D}^\pm (\lambda) E_{\mp \beta} \right\rangle \\
&= i  \tau_\beta^\pm (\lambda,t) \exp(\pm \vec{d}^\pm (\lambda),\beta) ,\\
\left.  i\left\langle  \left( \hat{\chi}^{\pm,\prime} (Q - \lambda J) \chi^{\pm,\prime}, E_{\pm\beta} \right\rangle \right) \right|_{x=-\infty}^\infty
&=-i \left\langle \hat{T}^\mp \mathcal{J}(\lambda))T^\mp (\lambda) E_{\pm \beta} \right\rangle \\
&= -i  \rho_\beta^\pm (\lambda,t).
\end{split}\end{equation}
The r.h.side of (\ref{eq:11.1}) provides skew-scalar product between $\ad_J^{-1}$ and the corresponding `squared solutions` of $L$:
\begin{equation}\label{eq:epmbe}\begin{split}
 e_{\pm \beta}^{\pm} =  \pi_J \chi^\pm E_{\pm \beta} \hat{\chi}^\pm            , \qquad
 e_{\pm \beta}^{\pm,\prime} = \pi_J \chi^{\pm,\prime} E_{\pm \beta} \hat{\chi}^{\pm,\prime}.
\end{split}\end{equation}
Indeed, from eqs. (\ref{eq:9.6}) we get the following relations:
\begin{equation}\label{eq:9.5'}\begin{aligned}
i \rho_\alpha^\pm (\lambda,t) e^{\pm (\vec{d}^\pm, \alpha)} & = \biglb \ad_J^{-1}Q_x, e^\pm_{\pm \alpha}(x,\lambda)\bigrb, &\quad
\rho_\alpha^\pm (\lambda,t) &= \left\langle \mathcal{J}(\lambda), T^\mp E_{\pm \alpha} \hat{T}^\mp (\lambda) \right\rangle, \\
 i \tau_\alpha^\pm (\lambda,t)  &= -\biglb \ad_J^{-1} Q_x, e_{\mp \alpha}^{\pm} \bigrb, &\quad  \tau_\alpha^\pm (\lambda,t)  &=
\left\langle \mathcal{J}(\lambda), S^\pm E_{\mp \alpha} \hat{S}^\pm (\lambda) \right\rangle,
\end{aligned}\end{equation}
and
\begin{equation}\label{eq:9.6'}\begin{aligned}
\rho_\alpha^\pm (\lambda,t) & = \biglb \ad_J^{-1}Q_x, e^{\pm,\prime}_{\pm \alpha}(x,\lambda)\bigrb, &\quad
\rho_\alpha^\pm (\lambda,t) &= \left\langle \mathcal{J}(\lambda), T^\mp E_{\pm \alpha} \hat{T}^\mp (\lambda) \right\rangle, \\
  \tau_\alpha^\pm (\lambda,t) e^{\mp (\vec{d}^\pm, \alpha)}   &= -\biglb \ad_J^{-1} Q_x, e_{\mp \alpha}^{\pm,\prime} \bigrb, &\quad  \tau_\alpha^\pm (\lambda,t)  &=
\left\langle \mathcal{J}(\lambda), S^\pm E_{\mp \alpha} \hat{S}^\pm (\lambda) \right\rangle.
\end{aligned}\end{equation}

\subsection{Wronskian relations for $\delta Q$}
We put restrictions on $\delta Q$ so that they preserve the characteristic polynomial of $U_\pm$, i.e.
\begin{equation}\label{eq:13.1}\begin{split}
 Q_\pm + \delta Q_\pm - \lambda J = - \widehat{\tilde{u}}_{0,\pm} \mathcal{J}(\lambda) \tilde{u}_{0,\pm}.
\end{split}\end{equation}
In fact we need $\delta Q \to 0$ for $x\to \pm \infty$.

For the Wronskian relations we need also the equation for $\delta \chi$:
\begin{equation}\label{eq:13.3a}\begin{split}
 i \frac{\partial \delta \chi}{ \partial x } + (Q-\lambda J) \delta \chi (x,t,\lambda) + \delta Q \chi(x,t,\lambda)=0.
\end{split}\end{equation}
Then we consider
\begin{equation}\label{eq:13.3}\begin{split}
 \left. i \hat{\chi} \delta \chi (x,t,\lambda) \right|_{-\infty}^\infty & = i \int_{-\infty}^{\infty} dx \; \left( \hat{\chi}_x \delta \chi
 +\hat{\chi} (\delta \chi)_x \right) \\
 &=  - \int_{-\infty}^{\infty} dx \; \hat{\chi} \delta Q \chi(x,t,\lambda).
\end{split}\end{equation}
Multiplying eq. (\ref{eq:13.3}) on the right by $E_\alpha$ and taking the Killing form we find:
\begin{equation}\label{eq:13.4}\begin{split}
  \left. \tr \left( i \hat{\chi} \delta \chi (x,t,\lambda) E_\alpha \right) \right|_{-\infty}^\infty & =
 -  \int_{-\infty}^{\infty} dx \; \tr \left( \hat{\chi} \delta Q \chi(x,t,\lambda) E_\alpha \right) \\
  &= \biglb \ad_J^{-1} \delta Q , e_\alpha^\pm (x,\lambda) \bigrb.
\end{split}\end{equation}
Thus from eqs. (\ref{eq:13.4}) we get the following relations:
\begin{equation}\label{eq:12.5'}\begin{aligned}
 \delta \rho_\alpha^\pm (\lambda,t) e^{\pm (\vec{d}^\pm, \alpha)} & = \biglb \ad_J^{-1} \delta Q, e^\pm_{\pm \alpha}(x,\lambda)\bigrb, &\quad
\delta \rho_\alpha^\pm (\lambda,t) &= \left\langle  \hat{T}^\mp (\lambda) \delta T^\mp, E_{\pm \alpha} \right\rangle, \\
 \delta \tau_\alpha^\pm (\lambda,t)  &= -\biglb \ad_J^{-1}\delta Q, e_{\mp \alpha}^{\pm}(x,\lambda) \bigrb, &\quad \delta \tau_\alpha^\pm (\lambda,t)  &=
\left\langle \hat{S}^\pm (\lambda) \delta S^\pm, E_{\mp \alpha}  \right\rangle,
\end{aligned}\end{equation}
and
\begin{equation}\label{eq:12.6'}\begin{aligned}
 \delta \rho_\alpha^\pm (\lambda,t) & = \biglb \ad_J^{-1} \delta Q, e^{\pm,\prime}_{\pm \alpha}(x,\lambda)\bigrb, &\quad
\delta \rho_\alpha^\pm (\lambda,t) &= \left\langle  \hat{T}^\mp (\lambda) \delta T^\mp , E_{\pm \alpha}  \right\rangle, \\
 \delta \tau_\alpha^\pm (\lambda,t) e^{\pm (\vec{d}^\pm, \alpha)}   &= -\biglb \ad_J^{-1} \delta Q, e_{\mp \alpha}^{\pm,\prime} (x,\lambda)\bigrb, &\quad
\delta \tau_\alpha^\pm (\lambda,t)  &= \left\langle \hat{S}^\pm (\lambda) \delta S^\pm , E_{\mp \alpha}  \right\rangle.
\end{aligned}\end{equation}

\subsection{The $t$-dependence of the scattering matrix}

Here we need the second operator of the Lax pair:
\begin{equation}\label{eq:13.2'}\begin{split}
 M\phi &\equiv \frac{\partial \phi}{ \partial t} + \left( \sum_{k=0}^{N-2}V_k (x,t)\lambda^k + \lambda^{N-1} (Q- \lambda J)\right) \phi(x,t,\lambda)=0.
\end{split}\end{equation}
We will also need the limits for $x\to \pm \infty$, namely:
\begin{equation}\label{eq:13.2''}\begin{split}
 \lim_{x\to \pm \infty} V_k(x,t) &= 0, \qquad k = 1,2,\dots ,N-2; \\
 M_{\rm \pm}\phi_{\rm \pm} & = i \frac{\partial \phi_{\rm \pm}}{ \partial t} +  \lambda^{N-1} (Q_\pm - \lambda J) \phi_{\rm \pm}; \\
 \phi_{ -} &= \hat{u}_{0,-} e^{-i \mathcal{J}x}, \qquad   \phi_{ +} = \hat{u}_{0,+} e^{-i \mathcal{J}x} T(\lambda,t).
\end{split}\end{equation}
In fact the compatibility condition requires thet
\begin{equation}\label{eq:13.3c}\begin{split}
 M \phi(x,t,\lambda) = \phi(x,t,\lambda) C(\lambda).
\end{split}\end{equation}
Consider first the limit of (\ref{eq:13.3c}) for $x\to -\infty$. We get:
\begin{equation}\label{eq:Mm}\begin{split}
 M_{-} \phi_{-} = \left( i \frac{\partial }{ \partial t} +\lambda^{N-1}(Q_- -\lambda J) \right) \hat{u}_{0,-}e^{-i \mathcal{J}x}
 = \hat{u}_{0,-}e^{-i \mathcal{J}x} C(\lambda),
\end{split}\end{equation}
i.e.
$
 C(\lambda) = -\lambda^{N-1} \mathcal{J}(\lambda)$.
Next consider the limit of (\ref{eq:13.3c}) for $x\to \infty$. We have:
\begin{equation}\label{eq:Tt}\begin{split}
 M_{+} \phi_{+} &= \left( i \frac{\partial }{ \partial t} +\lambda^{N-1}(Q_- -\lambda J) \right)\hat{u}_{0,+}e^{-i \mathcal{J}x} T(\lambda, t) = \hat{u}_{0,+}e^{-i \mathcal{J}x} T(\lambda,t) C(\lambda).
\end{split}\end{equation}
Thus:
\begin{equation}\label{eq:u0p}\begin{split}
 \hat{u}_{0,+} e^{-i \mathcal{J}x} \left( i \frac{\partial T}{ \partial t} - \lambda^{N-1} \mathcal{J} T(\lambda,t) \right)
 = - \hat{u}_{0,+} e^{-i \mathcal{J}x} T(\lambda,t) \lambda^{N-1} \mathcal{J} .
\end{split}\end{equation}
The final result is:
\begin{equation}\label{eq:Tt'}\begin{split}
 i \frac{\partial T}{ \partial t} = \lambda^{N-1} [\mathcal{J}, T(\lambda,t)].
\end{split}\end{equation}
It is easy to find also the $t$ dependence of the Gauss factors of $T(\lambda,t)$:
\begin{equation}\label{eq:Spmt}\begin{aligned}
 i \frac{\partial S^\pm}{ \partial t} &= \lambda^{N-1} [\mathcal{J}, S^\pm(\lambda,t)], \qquad
 i \frac{\partial T^\pm}{ \partial t} &= \lambda^{N-1} [\mathcal{J}, T^\pm(\lambda,t)] ,  \qquad i \frac{\partial D^\pm}{ \partial t} &= 0.
\end{aligned}\end{equation}
Now we assume that the variations of the scattering data in eqs. (\ref{eq:12.5'}) and (\ref{eq:12.6'}) is due to the
evolution of the scattering matrix along the NLEE defined by the $M$ operator (\ref{eq:13.2'}). Using (\ref{eq:Spmt}) we get:
\begin{equation}\label{eq:deltau}\begin{split}
 \delta \tau^\pm_{\alpha} &\rightarrow \left\langle \hat{S}^\pm \frac{\partial S^\pm}{ \partial t}, E_{\mp \alpha} \right\rangle \delta t
 = \lambda^{N-1} \delta t \left\langle \hat{S}^\pm  [ \mathcal{J}, S^\pm], E_{\mp \alpha} \right\rangle  = \lambda^{N-1} \delta t \tau^\pm _\alpha (\lambda), \\
\delta \rho^\pm_{\alpha} &\rightarrow \left\langle \hat{T}^\mp \frac{\partial T^\mp}{ \partial t}, E_{\pm \alpha} \right\rangle \delta t
 = \lambda^{N-1} \delta t \left\langle \hat{T}^\mp  [ \mathcal{J}, T^\mp], E_{\pm \alpha} \right\rangle = \lambda^{N-1} \delta t \tau^\pm _\alpha (\lambda).
\end{split}\end{equation}
Thus for the expansion coefficients of $Q_t$ we obtain:
\begin{equation}\label{eq:12.5c}\begin{aligned}
i \frac{\partial \rho_\alpha^\pm }{ \partial t} e^{\pm (\vec{d}^\pm, \alpha)} & = \biglb \ad_J^{-1}  Q_t, e^\pm_{\pm \alpha}(x,\lambda)\bigrb, &\quad
i \frac{\partial  \tau_\alpha^\pm }{ \partial t}  &= -\biglb \ad_J^{-1} Q_t, e_{\mp \alpha}^{\pm} \bigrb,
\end{aligned}\end{equation}
and
\begin{equation}\label{eq:12.6''}\begin{aligned}
i \frac{\partial \rho_\alpha^\pm }{ \partial t} & = \biglb \ad_J^{-1}  Q_t, e^{\pm,\prime}_{\pm \alpha}(x,\lambda)\bigrb, &\quad
i \frac{\partial  \tau_\alpha^\pm }{ \partial t} e^{\pm (\vec{d}^\pm, \alpha)}   &= -\biglb \ad_J^{-1} Q_t, e_{\mp \alpha}^{\pm,\prime} \bigrb.
\end{aligned}\end{equation}

\section{The $N$-wave hierarchy}\label{sec:4}

The dispersion laws of the $N$-wave and MNLS hierarchy may contain several $\sqrt{\lambda^2 - \rho_k^2}$.

As we mentioned above, the $x$-derivatives $Q_x$ of the potentials with CBC {\it vanish} for $x\to \pm \infty$. In other
words they are elements of linear spaces, and the `squared solutions` are proper tools for the GFT.

In addition we need the recursion operators which have the `squared solutions` as eigenfunctions.
Note that the potential $Q(x,t)$ and its derivative $Q_x$ are restricted to the image of the operator $\ad_J$. Thus the
recursion operator also must be restricted to the image of the operator $\ad_J$. This means that we have to split the
`squared solutions` as follows:
\begin{equation}\label{eq:split}\begin{split}
 e^\pm_\alpha (x,t,\lambda) = e^{\pm, \rm f}_\alpha (x,t,\lambda) + e^{\pm, \rm d}_\alpha (x,t,\lambda),
\end{split}\end{equation}
where $e^{\pm, \rm d}_\alpha (x,t,\lambda) \in \mathfrak{h}$, in other words is diagonal matrix.

Then the derivation follows the well known steps. First `squared solutions obviously satisfied the equation
\begin{equation}\label{eq:Le}\begin{split}
 i \frac{\partial e^\pm_\alpha}{ \partial x } + [Q(x,t) - \lambda J, e^{\pm}_\alpha (x,t,\lambda)] =0.
\end{split}\end{equation}
Insert the splitting (\ref{eq:split}) into (\ref{eq:Le}) and treat separately the diagonal and the off-diagonal
parts of $e^\pm_\alpha$. For the diagonal part we have:
\begin{equation}\label{eq:Led}\begin{split}
 i \frac{\partial e^{\pm,\rm d}_\alpha}{ \partial x } + [ Q(x,t),  e^{\pm, \rm f}_\alpha (x,t,\lambda)]^{\rm d} =0,
\end{split}\end{equation}
or
\begin{equation}\label{eq:Led0}\begin{split}
 i \frac{\partial e^{\pm,\rm d}_\alpha}{ \partial x } + \sum_{k=1}^{r} \left\langle
 [ Q(x,t)  e^{\pm, \rm f}_\alpha (x,t,\lambda)]^{\rm d}, H_k \right\rangle H_k =0,
\end{split}\end{equation}
where the Cartan generators $H_k$ are normalized by $\langle H_j, H_k \rangle = \delta_{jk}$.

Integrating once we find:
\begin{equation}\label{eq:Led1}\begin{split}
  e^{\pm, \rm d}_\alpha (x,t,\lambda) = i \int_{\epsilon \infty}^{x} \sum_{k=1}^{r} \left\langle
  [ Q(x,t) , e^{\pm, \rm f}_\alpha (x,t,\lambda)], H_k \right\rangle H_k + \lim_{x\to \epsilon \infty}
  e^{\pm, \rm d}_\alpha (x,t,\lambda),
\end{split}\end{equation}
where $\epsilon = \pm 1$. For the off-diagonal part we obtain:
\begin{equation}\label{eq:Lef}\begin{split}
i \frac{\partial e^{\pm,\rm f}_\alpha}{ \partial x } + [ Q(x,t),  e^{\pm, \rm f}_\alpha (x,t,\lambda)]^{\rm f}
+ [ Q(x,t),  e^{\pm, \rm d}_\alpha (x,t,\lambda)] = \lambda [J, e^{\pm, \rm f}_\alpha (x,t,\lambda)].
\end{split}\end{equation}
Now we apply $\ad_J^{-1}$ and obtain the formal expression for recursion operator $\Lambda_\epsilon$:
\begin{equation}\label{eq:Lae}\begin{split}
 \Lambda_\epsilon X &= \ad_J^{-1} \left( i \frac{\partial X}{ \partial x } + [ Q(x,t), X] ^{\rm f} + i \left[ Q(x,t),   \int_{\epsilon \infty}^{x} \sum_{k=1}^{r} H_k \left\langle   [ Q(x,t) , X], H_k \right\rangle  \right] \right),
\end{split}\end{equation}
where  $X \equiv X^{\rm f}$. Acting with it on the `squared solutions` and making use of (\ref{eq:Led1}) we find:
\begin{equation}\label{eq:Lae'}\begin{split}
\Lambda_\epsilon e^{\pm, \rm f}_\alpha (x,t,\lambda) = \lambda e^{\pm, \rm f}_\alpha (x,t,\lambda)  +
 \left[ Q(x,t),  \lim_{x\to \epsilon \infty}  e^{\pm, \rm d}_\alpha (x,t,\lambda)\right].
\end{split}\end{equation}
Therefore the `squared solutions` $e^{\pm, \rm f}_\alpha (x,t,\lambda)$ will be eigenfunctions of $\Lambda_\epsilon$ if and
only if 
\[
\lim_{x\to \epsilon \infty}  e^{\pm, \rm d}_\alpha (x,t,\lambda) =0.
\]
Some $e^{\pm, \rm f}_\alpha (x,t,\lambda)$ can not be an eigenfunction of $\Lambda_\epsilon$, but convenient linear combination of
them are eigenfunctions.

\subsection{Higher $N$-wave equations}

Here we will use an equivalent, but slightly different form of the Lax operator:
\begin{equation}\label{eq:Lt}\begin{split}
 \mathcal{L}\psi \equiv i \frac{\partial \psi}{ \partial x } + (Q(x,t) - \lambda J) \psi(x,t,\lambda) =0, \qquad Q\in \mathfrak{g}/\mathfrak{h}.
\end{split}\end{equation}
Indeed we may use a hierarchy of $M$ operators which are polynomials of $\lambda$ of any order $p$:
\begin{equation}\label{eq:Mp}\begin{split}
 \mathcal{M}^{(N)} &\equiv i \frac{\partial \psi}{ \partial t} + (V(x,t,\lambda) -\lambda^N J)\psi (x,t,\lambda) =0, \\
 V(x,t,\lambda) &= \sum_{k=1}^{N} V_k(x,t) \lambda^{N-k}.
\end{split}\end{equation}
Again we request that the compatibility condition:
\begin{equation}\label{eq:LMc}\begin{split}
[ \mathcal{L}, \mathcal{M}^{(N)} ]=0
\end{split}\end{equation}
holds identically with respect to $\lambda$. The leading term in $\lambda$ in (\ref{eq:LMc}) is $\lambda^{N+1} [J,J] =0$.
The next order terms are:
\begin{equation}\label{eq:LM2}\begin{aligned}
& \lambda^p &\colon \quad & -[Q,J] -[J,V_1]  =0, &\quad  &V_1 = Q, \\
& \lambda^{N-1} &\colon \quad & [Q,V_1] + iV_{1,x} -[J,V_2]  =0, &\quad  &V_2 = i \ad_J^{-1} Q_x, \\
& \lambda^{N-k} &\colon \quad & iV_{k,x} +[Q,V_k]  -[J,V_{k+1}] =0, &\quad & k = 2, 3, \dots, N-1;  \\
&\lambda^0 & \colon \quad  &- i Q_t + i V_{N,x} + [Q, V_N]=0.
\end{aligned}\end{equation}
The first two relations in (\ref{eq:LM2}) allowed us to determine easily the first two coefficients
$V_1$ and $V_2$ in $\mathcal{M}$. The rest of these equations we may interprete as recurrent relations,
which will allow us to express all $V_k$ in terms of $Q$ and its $x$-derivatives. In order to demonstrate this we
need to split each $V_k$ into diagonal and off-diagonal parts:
\begin{equation}\label{eq:Vkp}\begin{split}
 V_k(x,t) =  V_k^{\rm d}(x,t) + V_k^{\rm f}(x,t), \qquad V_k^{\rm d}(x,t) = \sum_{j=1}^{r} w_{k,j}(x,t) H_j,
\end{split}\end{equation}
where $w_{k,j} = \langle V_k(x,t), H_j\rangle$. The Cartan generators $H_j$ are normalized by $\langle H_j, H_k\rangle = \delta_{jk}$ and $V_k^{\rm f}(x,t) \in \mathfrak{g}/\mathfrak{h}$.
Then the diagonal part of the $p-k$ equation in (\ref{eq:LM2}) gives:
\begin{equation}\label{eq:k0}\begin{split}
 iV_{k,x}^{\rm d} +[Q,V_k^{\rm f}]^{\rm d} =0.
\end{split}\end{equation}
Integrating on $x$ we find:
\begin{equation}\label{eq:wkj}\begin{split}
 w_{k,j} (x,t) = i \int_{x}^{\epsilon \infty} \langle H_j, [Q,V_k^{\rm f}] \rangle + \lim_{x\to \epsilon} V_{k}^{\rm d}.
\end{split}\end{equation}
The off-diagonal part of the same equation gives:
\begin{equation}\label{eq:Vkf}\begin{split}
& iV^{\rm f}_{k,x} +[Q,V_k]^{\rm f} + [Q, V_k^{\rm d}  =[J,V^{\rm f}_{k+1}] , \qquad \mbox{i.e.}  \qquad  V_{k+1}^{\rm f} = \Lambda_\epsilon V_{k}^{\rm f} ,
\end{split}\end{equation}
where $\Lambda_\epsilon$ is the recursion operator (\ref{eq:Lae}).
Thus we have outlined the generalization of AKNS ideas \cite{AKNS,vsg82} for homogeneous spaces. The MNLS equation (\ref{eq:mnls})
is a consequence from the above consideration for $p=2$. It is well known \cite{ForKu,Contemp} that  $(\Lambda_\epsilon)^k Q$
is local, ie. it is expressed in terms of $Q$ and its $x$-derivatives.

Finally, the corresponding NLEE, i.e. the last equation of (\ref{eq:LM2}) becomes:
\begin{equation}\label{eq:HNwa}\begin{split}
 i \frac{\partial Q}{ \partial t} + \Lambda_\epsilon^{N-1} \ad_J^{-1} \frac{\partial Q}{ \partial x } =0.
\end{split}\end{equation}
Comparing these results with the ones in Subsection 3.3 we conclude that if $Q(x,t)$ satisfies the above NLEE then the
scattering data will satisfy eq. (\ref{eq:Spmt}). In particular, for $N=2$ eq. (\ref{eq:HNwa}) becomes the NLS eq.

\section{Conclusions}
Here we analyzed the spectral properties of the $N$-wave hierarchy with CBC. We paid most attention to
the higher $N$-wave equations whose $M$-operators are polynomial of order $\lambda^N$ with $N\geq 2$. For these equations we
determined the dispersion laws which have rather nontrivial character: $\lambda^{N-1} \mathcal{J}(\lambda)$ which contain
at least two $\pm \sqrt{\lambda^2 a_k^2 -\rho_k^2}$, $k=1,2$. It will be important to physical phenomena that are compatible
with such dispersion laws.  Similar analysis can be done for  $N$-wave resonant interaction models with more general boundary conditions.

The next natural step would be to demonstrate that the Inverse scattering problems applied to this class of Lax
operators also can be understood as generalized Fourier transform, see \cite{gvy08,GeYa,GeYa*13,GeYa,SIAM,Contemp}. Such results
will be based on the completeness relations for the `squared solutions` $e^\pm_\alpha (x,t,\lambda)$.

It is also an open problem to  derive the soliton solutions of \eqref{eq:N-waves} in the case of constant boundary conditions (the so-called "dark solitons"),  various types of dark-dark and dark-bright soliton solutions \cite{ABP2} by modifying the dressing Zakharov-Shabat method \cite{ZMNP}, or by using the Darboux transformation method \cite{Deg}.

Another open problem is to study the behavior of the scattering data at the end-points of the continuous spectrum in the complex $\lambda$-plane;
this requires generalization of the the method developed in  \cite{FaTa}.


\section*{Acknowledgements} We are grateful to Dr. T. Valchev for the useful discussions. This work is supported by the Bulgarian National Science Fund, grant KP-06N42-2.

\appendix

\section{Lax pairs for NLEE on homogeneous spaces}\label{sec:A1}

 If we choose the $M$-operator to be a quadratic polynomial in $\lambda$
\[
M_2(\lambda)=\left( i{d \over dt} + V_0(x,t)(x,t)+
\lambda V_1(x,t)+\lambda^2V_2(x,t)\right) \Psi(x,t,\lambda ) = 0,
\]
then the AKNS scheme will allow us to determine recursively all its coefficients in terns of $Q$ and the adjoint action operator ${\rm ad}_J$. This will give the second member of the integrable hierarchy associated to $L$ of the form \eqref{eq:L}:
\begin{equation}\label{eq:mnls}\begin{split}
 i \frac{\partial Q}{ \partial t} + \ad_J^{-1}\frac{\partial^2 Q}{ \partial x^2} + \left[ \ad_J^{-1} Q_x
,Q\right]  + [w(x,t), Q(x,t)] =0.
\end{split}\end{equation}
This is a multicomponent nonlinear Schr\"odinger system related to Homogeneous spaces \cite{ForKu}. The construction  and the structure of its Lax pair are compatible with the structure of the homogeneous spaces, see \cite{Helg}.

As an example, if we take the potential \eqref{eq:sp4u}, then we can get the second member of the $4$-wave hierarchy related to $sp(4,{\Bbb C})$:
\[
\ad_j^{-1} Q(x,t) = \left(\begin{array}{cccc} 0 & \displaystyle{\frac{q_4 }{a_1-a_2}} & \frac{q_3 }{a_1+a_2} & \frac{q_1 }{2a_1} \\
- \frac{s_4 }{a_1-a_2} & 0 &  \frac{q_2 }{2a_2} & -\frac{q_3 }{a_1+a_2} \\ -\frac{s_3 }{a_1+a_2}  & -\frac{2_2 }{2a_2}  & 0 & \frac{q_4 }{a_1-a_2} \\
  -\frac{s_1 }{2a_1} & \frac{s_3 }{a_1+a_2} & -\frac{s_4 }{a_1-a_2} &  0   \end{array}\right),
\]
\begin{equation}\label{eq:mnls1}\begin{split}
 w_1 (x,t) &= -\frac{2 q_4 s_4 }{a_1-a_2} -\frac{2 q_3 s_3 }{a_1+a_2} -\frac{ q_1 s_1 }{a_1}   , \\
w_2(x,t) &= \frac{2 q_4 s_4 }{a_1-a_2} -\frac{2 q_3 s_3 }{a_1+a_2} -\frac{ q_2 s_2 }{a_2}.
\end{split}\end{equation}
where $w = \diag( w_1, w_2, -w_2, -w_1) $ and $s_k = - q_k^*$.

\end{document}